\documentclass[letterpaper]{article} 
\usepackage{aaai24}  
\usepackage{times}  
\usepackage{helvet}  
\usepackage{courier}  
\usepackage[hyphens]{url}  
\usepackage{graphicx} 
\usepackage{natbib}  
\usepackage{caption} 
\usepackage{algorithm}
\usepackage{algorithmic}
\usepackage{newfloat}
\usepackage{listings}
\DeclareCaptionStyle{ruled}{labelfont=normalfont,labelsep=colon,strut=off} 
\floatstyle{ruled}
\newfloat{listing}{tb}{lst}{}
\floatname{listing}{Listing}
\title{Friendly Attacks to Improve Channel Coding Reliability}
\author {
    Anastasiia Kurmukova and Deniz Gunduz
}
\affiliations {
    Department of Electrical and Electronic Engineering, 
    Imperial College London, London, UK \\
    a.kurmukova22@imperial.ac.uk,  d.gunduz@imperial.ac.uk
}

\usepackage{url}            
\usepackage{booktabs}       
\usepackage{amsfonts}       
\usepackage{nicefrac}       
\usepackage{microtype}      
\usepackage{tikz}
\usetikzlibrary{shapes.geometric, arrows}
\tikzstyle{startstop} = [rectangle, rounded corners, minimum width=1cm, minimum height=1cm,text centered, draw=black, fill=red!30]
\tikzstyle{modulators} = [rectangle, rounded corners, minimum width=1cm, minimum height=1cm,text centered, draw=black, fill=blue!30]
\tikzstyle{arrow} = [thick,->,>=stealth]
\usepackage{amsmath}
\DeclareMathOperator*{\argmax}{arg\,max}
\DeclareMathOperator*{\argmin}{arg\,min}
\DeclareMathOperator*{\sign}{sign}
\DeclareMathOperator*{\BLER}{BLER}
\DeclareMathOperator*{\BER}{BER}

\begin{document}

\maketitle

\begin{abstract}
    This paper introduces a novel approach called "friendly attack" aimed at enhancing the performance of error correction channel codes. Inspired by the concept of adversarial attacks, our method leverages the idea of introducing slight perturbations to the neural network input, resulting in a substantial impact on the network's performance. By introducing small perturbations to fixed-point modulated codewords before transmission, we effectively improve the decoder's performance without violating the input power constraint. The perturbation design is accomplished by a modified iterative fast gradient method. This study investigates various decoder architectures suitable for computing gradients to obtain the desired perturbations. Specifically, we consider belief propagation (BP) for LDPC codes; the error correcting code transformer, BP and neural BP (NBP) for polar codes, and neural BCJR for convolutional codes. We demonstrate that the proposed friendly attack method can improve the reliability across different channels, modulations, codes, and decoders. This method allows us to increase the reliability of communication with a legacy receiver by simply modifying the transmitted codeword appropriately. 

\end{abstract}

\section{Introduction}

Channel coding plays a crucial role in modern communication systems, ensuring reliable information transmission over noisy channels. Over the past 70 years, remarkable progress has fueled the continuous development of this field, yielding robust and efficient communication protocols that employ a wide range of hand-crafted codes. Contemporary standards such as Long Term Evolution (LTE) and 5G \cite{3gpp} have been significantly influenced by the development of near-optimal channel codes, including linear block codes such as polar codes \cite{polar} and low density parity check (LDPC) codes \cite{LDPC}, alongside linear codes with memory, such as turbo codes \cite{turbo} based on convolutional codes \cite{elias}. These codes demonstrate capacity-approaching performance \cite{shannon_capacity} on additive white Gaussian noise (AWGN) channels; however, their optimal decoding is still a challenging problem.

Despite progress, a gap persists between achievability bounds based on random coding \cite{bounds_polyanskiy} and the performance of hand-crafted codes in the short block length regime. Maximum likelihood decoding of linear codes is known to be NP-hard. Iterative algorithms, such as belief propagation (BP) operating over the code Tanner graph, present an efficient alternative. However, BP decoding demonstrates near-capacity performance primarily for LDPC codes with large block lengths \cite{LDPC_shannon}. For polar codes, BP decoding performance lags behind successive cancellation (SC) decoding \cite{polar} attributed to a great number of short cycles in their Tanner graphs.

Decoding of channel codes can be considered as a classification problem, where the aim is to infer the transmitted codeword from its noisy version. Therefore, the recent success of deep learning methods for a wide range of classification problems and their ability to capture complex patterns have motivated exploring their applications to channel decoding. Neural BP (NBP), which introduces trainable weights to conventional BP graphs, has gained popularity due to its simplicity and efficiency \cite{neural_bp1, neural_bp2, neural_minsum}. Neural BP decoding of polar code has enhanced the code performance \cite{neural_bp_polar, neural_bp_polar_crc}. Recently, a significant improvement is achieved in NBP decoding with two-stage decimation \cite{neural_bp_impr}.

There are other model-based neural decoding approaches that leverage existing non-neural decoders' structure to enhance performance with trainable architectures. Notably, neural successive cancellation for polar codes \cite{neural_sc}, neural BCJR (NBCJR) for convolutional codes \cite{neuralbcjr}, KO-codes for Reed-Muller codes \cite{KOcodes} have shown promising results. Conversely, model-free deep learning decoding approaches, like \cite{on_deeplearning_based_cc} and \cite{intro_to_ml_cc}, which do not use any existing decoding algorithm as a baseline, are limited to short block lengths ($k < 16$) due to the curse of dimensionality. Syndrome-based schemes \cite{syndrom_based} and the error correction code transformer (ECCT) \cite{ECCT} have recently achieved performance improvements by embedding the code structure through the syndromes into the learnable decoding architecture.

The improvement in transmission performance extends beyond the enhancement of channel decoders. In real-world transmission, fixed-constellation modulation is usually employed, involving a uniform mapping of a bit sequence (or a codeword) into a sequence of symbols from a finite set of constellation points. Examples of such modulations include commonly used binary phase shift keying (BPSK) modulation (with constellation $\{1, -1\}$) and higher order modulations like quadrature phase shift keying (QPSK), $2^n$ quadrature amplitude modulation (QAM). It was shown that the traditional choice of the constellation may not be optimal for various communication systems, therefore the modulation performance can be enhanced by the technique called signal shaping \cite{prob_geom_shaping}. There are two prevalent approaches allowing the selection of more suitable modulations for transmission: geometrical shaping and probabilistic shaping. The first method involves locating optimal transmission constellation points according to specific criteria, while the second method adjusts the probabilities of transmitted symbols to maximize the mutual information between the channel input and the output. 

Learning based techniques employing an autoencoder architecture have also been used for geometric shaping \cite{intro_to_ml_cc, Jones:ECOC:19, Gumus:OFC:20} as well as joint geometric and probabilistic shaping \cite{Aref:OFC:22}. While they provide considerable gains, they rely on modifying both the encoder and decoder architectures, which may not be viable in some practical systems. In contrast, the goal in this work is to increase the reliability of communication with a fixed oblivious receiver, by only modifying the transmitted signals. This can allow, for example, higher reliability in the downlink of cellular communication networks or digital radio/TV broadcasting systems without upgrading mobile devices. 

We treat the modifications to the transmitted symbols as perturbations to the modulated codewords. The goal is to identify the ideal perturbations that are more likely to result in correct decoding given a fixed decoding rule, e.g., fixed decoding boundaries. This can be considered as the converse of adversarial attacks on neural networks \cite{fgsm}. We propose a new concept of `friendly attacks' for channel encoding, where the goal is to find a perturbation to the modulated codeword that improves the decoding performance without violating the average power constraint at the transmitter.

The underlying idea of a `friendly attack' is as follows. The space of the codewords is divided by the decoder into decision regions. Modulated codewords may be suboptimal for these decision regions, particularly when the decoding algorithm is suboptimal. The optimal codeword locations should be as far from the decision boundaries as possible. Hence, a friendly attack seeks a perturbation vector (or, an attack vector) that shifts the modulated codeword towards the center of the corresponding decision region.

We design the attack vector by finding gradients during noisy channel transmission and subsequent decoding, akin to white-box adversarial attacks \cite{fgsm, ifgsm, scale_attack}. The computed attack vector is added to the modulated codeword before transmission. We highlight the differences of friendly attacks compared to conventional adversarial attacks on neural networks: The most obvious difference is that our goal is to improve the quality of decisions made by the receiver, while adversarial attacks try to fool the neural network. Moreover, in adversarial attacks, we determine a specific attack for a specific input that deterministically results in a wrong decision. In the case of channel decoding, codewords are decoded correctly in most cases, while errors happen rarely, when the random channel noise moves the received codeword to the wrong decision region. Hence, our goal is to modify the channel input to increase the correct decoding probability. Finally, in adversarial attacks, perturbations are bounded in order to limit their perception by humans. Here, perturbations are bounded to guarantee that the average power of the transmitted codeword remains the same after its application. We also would like to emphasize that, thanks to the linearity of the code, there is no need to find a distinct specific perturbation for each codeword. It is sufficient to find a perturbation for the all-zero codeword, which can then be applied to any codeword resulting in a similar improvement to the decoding performance. 

The friendly attack approach to channel coding proposed here is applicable to any code, and any neural or differentiable decoding algorithm. We provide the results of successful friendly attacks for following scenarios: LDPC code $n=64, k=32$ for BP decoder with BPSK and $4$-QAM modulations; polar code $n=64, k=32$ for BP, NBP and ECCT decoders with BPSK and $4$-QAM modulations and also for fading channel with BPSK; long block length polar code with $n=512, k=256$ for BP and NBP with BPSK modulation; convolutional code $k = 100, R=\frac{1}{2}$ with NBCJR decoder over AWGN and bursty channels. We show that the proposed friendly attack can improve the decoding performance significantly for a suboptimal decoder in the high SNR regime. Our results show that thoroughly designed attack vector enhancing BP decoding for a chosen code can also improve the performance of a BP decoder with trainable weights for the same code and the same number of iterations. Moreover, for the considered codes the improvement by friendly attack for the BER of BP decoding with several iterations is greater than improvement achieved by adding trainable weights to BP decoding (NBP). The idea of perturbing the modulated codeword with an attack vector was also proposed in \cite{Hameed:TIFS:21}, but there the goal is to prevent an eavesdropper trying to detect the modulation scheme of the transmitted waveform; hence, it follows the conventional adversarial attack framework.

\section{Background}

We use the following notations for vectors: $\textbf{x}^n = \{x_1, x_2, \dots, x_n\}$, $\textbf{x}_i^j = \{x_i, x_{i+1}, \dots, x_j\}$; a bold notation $\textbf{x}^n$ for a vector with fixed values and $X^n$ for the corresponding vector of random variables.

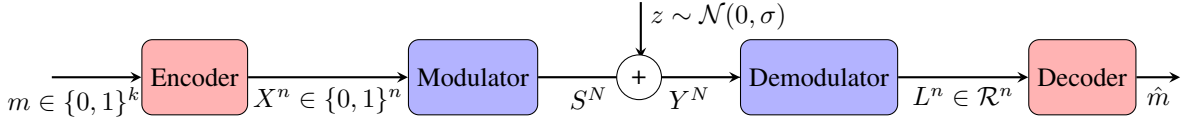
\begin{figure*}[h]
    \centering
    \begin{tikzpicture}[node distance=3cm]
        \path (0,0) node(m) {}
        (1.9,0) node[startstop](enc) {Encoder};
        (7.83,1) node(noise) {}
        (16,0) node(end) {}
        \draw [arrow] (7.83,1) -- node[near start,right] {$z \sim \mathcal{N}(0,\sigma)$} (7.83,0.25);

        \draw [arrow] (0,0) -- node[near start,below] {$m\in\{0,1\}^k$} (enc);
        \node (mod) [modulators, right of=enc, xshift=0.7cm] {Modulator};
        \node (demod) [modulators, right of= mod, xshift=1.6cm] {Demodulator};
        \node (decoder) [startstop, right of=demod, xshift=0.5cm] {Decoder};
        \draw [arrow] (enc) -- node[below] {$X^n\in\{0,1\}^n$} (mod);
        \draw [arrow] (mod) -- node[near start,below] {${S}^N$} node[near end,below] {${Y}^N$}(demod);
        \draw [arrow] (demod) -- node[below] {$L^n\in \mathcal{R}^n$} (decoder);
        \draw [arrow] (decoder) -- node[below] {$\hat{m}$} (15, 0);
        \draw (mod) edge node[draw,fill=white,circle]{+} (demod);
    
    \end{tikzpicture}
    \caption{The studied coded communication scenario with AWGN channel.}
    \label{fig:scheme}
\end{figure*}

\subsection{Communication Scheme}

Communication scheme, as shown in Fig. \ref{fig:scheme}, utilizes a channel code for encoding information at the transmitter and decoding information from a noisy sequence at the receiver. For encoding and decoding we assume traditional transmission setting with a linear $(n, k)$ code $\mathcal{C}$ over a binary field $\mathbb{F} = \{0, 1\}$. The sender has a $k$-bit message $m\in\{0,1\}^k$, which is encoded into a codeword $X^n$ from the code space $\mathcal{C}$. The modulator then maps the sequence of codeword bits into a sequence of symbols $\textbf{s}^N$ from a chosen discrete constellation, and the modulated codeword is transmitted over the physical channel. For instance, BPSK modulation has the constellation $\{1, -1\}$, and the modulation can be written as:
\begin{equation}
    S = 1 - 2X \ . \nonumber
\end{equation}
We will also consider modulations with a higher constellation order such as $4$-QAM. Here we suppose that the communication scheme must follow the average power constraint for the codeword. Every modulated codeword $\textbf{s}^N$ must satisfy average power constraint $P$: ${||\textbf{s}^N||}_2^2 \leq NP$. 

We consider an additive white Gaussian noise (AWGN) channel given as follows:
\begin{equation}
    Y^{N} = S^N + Z^N,  \nonumber
\end{equation}
where $Z^N$ is a sequence of independent and identically distributed (i.i.d.) Gaussian random variables $Z_i \sim \mathcal{N}(0, \sigma^2)$. We will also consider a Rayleigh fading channel \cite{rayleigh}, where
\begin{equation}
    Y^{N} = G^N S^N + Z^N,  \nonumber
\end{equation}
where channel gains $G^N$ are i.i.d. random variables from Rayleigh distribution, and $Z^N$ are defined in the same way as in AWGN channel. If the receiver knows channel gains $G^N$, then it is the case of the ideal side information (SI), otherwise there is no side information.

We also consider an AWGN channel with bursty noise, which approximates interference in wireless channels:
\begin{equation}
    Y^{N} = S^N + Z^N + W^N,  \nonumber
\end{equation}
where $Z_i \sim \mathcal{N}(0, \sigma^2)$, and a component of bursty noise $W_i \sim \mathcal{N}(0, \sigma^2_b)$ with probability $\rho$ and  $W_i = 0$ with probability $1 - \rho$. The bursty noise parameter $\sigma_b^2$ is usually taken quite large to have noticeable influence on the performance.

The decoder receives $Y^{N}$ and computes the log likelihood ratios (LLRs) $L^n$ during demodulation. For each bit in a codeword $L$ is computed as:
\begin{equation}
    L = \log \frac{\Pr(X=0|Y^{N})}{\Pr(X=1|Y^{N})}\ . \nonumber
\end{equation}
Then the LLRs vector $L^n$ is passed to the decoder which recovers a message $\hat{m} \in\{0,1\}^k$.

\subsection{Channel Coding}

We limit our consideration of the hand-crafted codes to binary linear codes. The encoding process for a binary linear block code $\mathcal{C}$ can be given by multiplication over the binary field $\mathcal{F}$ by the binary generator matrix $G$ of size $k \times n$: $X^n = mG$, $m \in \{0,1\}^k$. The code rate is defined as $R = \frac{k}{n}$.

The aim of the decoding process is to recover the transmitted message $m$ by exploiting the code structure. The hard-output decoders return a hard decision $\hat{m}$ that is an estimate of transmitted message $m$. Successive cancellation (SC) \cite{polar} and successive cancellation list (SCL) \cite{scl_polar} for polar codes are hard-output decoders which show near optimal performance though have a low decoding throughput because of sequential decoding. Soft-output decoders return a soft estimate ${\hat{X}}^n \in \mathcal{R}^n$ instead of a hard decision $\hat{m}$. The sign of each position in the soft output ${\hat{X}}^n$ is an estimate of the codeword bit, and the hard decision $\hat{m}$ can be then obtained from the recovered codeword. The absolute values of ${\hat{X}}^n$ can be interpreted as a decoder's confidence for each bit. This is especially useful for decoding the concatenation of codes. One of the common examples of the soft-output decoder is an iterative BP algorithm which we describe further.

Commonly used metrics that measure decoding performance are the BLER (block error rate) defined as:
    \begin{equation}
        \BLER = \Pr\left(\hat{m} \neq m\right),
    \end{equation}
and the BER (bit error rate) defined as:
    \begin{equation}
        \BER = \frac{1}{k}\sum_{i=1}^{k}\Pr\left(\hat{m}_i \neq m_i\right).
    \end{equation}

\subsection{BP Decoding}

For decoding a binary linear block code $\mathcal{C}$ let the parity check matrix $H$ of size $(n-k) \times n$ be such that $GH^T = 0$ over the binary field $\mathcal{F}$. Thus, each codeword $X^n \in \mathcal{C}$ satisfies $HX^n = 0$. BP is a message passing algorithm commonly used for decoding of linear block codes. BP decoding is performed over the Tanner graph that is a bipartite graphical representation of the parity check matrix $H$. BP iteratively exchanges messages between variable nodes (representing codeword symbols) and check nodes (representing parity constraints). 
As BP decoding performance for polar codes suffers from short cycles in the factor graph, recent research has improved BP decoding through factor graph permutations \cite{bp_polar_permuted1, bp_polar_permuted2}, and early stop aided by additional CRC \cite{bp_polar_earlystop}. 

\begin{figure*}[h]
    \centering
    \begin{tikzpicture}[node distance=2.6cm]
        \path (0,0) node(m) {}
        (1.5,0) node[startstop](enc) {Encoder};
        (6.2,1) node(attack) {}
        (7.5,1) node(noise) {}
        (14,0) node(end) {}
        \draw [arrow] (6.2,1) -- node[near start,right] {$a^N$} (6.2,0.4);
        \draw [arrow] (7.5,1) -- node[near start,right] {$z \sim \mathcal{N}(0,\sigma)$} (7.5,0.25);

        \draw [arrow] (0,0) -- node[near start,below] {$m$} (enc);
        \node (mod) [modulators, right of=enc] {Modulator};
        \node (plus1) [draw, rectangle,fill=white, right of=mod,text centered, text width=0.7cm, xshift=-0.5cm] {+ \& norm};
        \node (plus2) [draw, circle,fill=white, right of=plus1, xshift=-1.3cm] {+};
        \node (demod) [modulators, right of=plus2, xshift=-0.5cm] {Demodulator};
        \node (decoder) [startstop, right of=demod, xshift=0.3cm] {Decoder};
        \draw [arrow] (enc) -- node[below] {$X^n$} (mod);

        \draw [arrow] (mod) -- node[below] {${S}^N$} (plus1);
        \draw [arrow] (plus1) -- node[below] {} (plus2);
        \draw [arrow] (plus2) -- node[below] {${Y}^N$} (demod);
        \draw [arrow] (demod) -- node[below] {$L^n$} (decoder);
        \draw [arrow] (decoder) -- node[below] {$\hat{m}$} (14, 0);

        \draw[draw=blue] (6.85,-0.6) rectangle ++(7.4,1.7);
        \path (9.85,1.7) --  (10.4,1.7) node[midway,below] {$f$};

    \end{tikzpicture}
    \caption{The communication scheme with a friendly attack $a^N$.}
    \label{fig:scheme_attack}
\end{figure*}
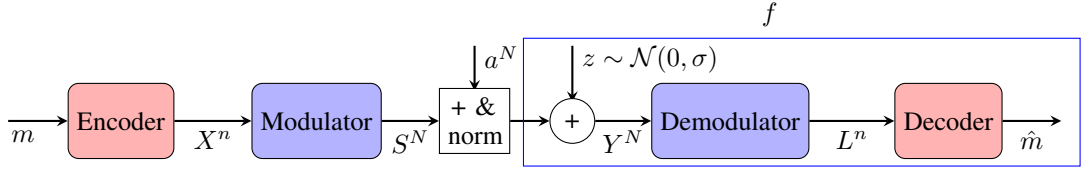

\subsection{Neural Decoding}

Both model-free and model-based neural approaches for channel decoding are topics of active research. Adding trainable weights to node updates in BP decoding resulted in a popular model-based neural decoding called NBP decoding \cite{neural_bp1, neural_bp2, neural_minsum}. One of the main challenges in neural decoding remains the choice of the loss function. Binary cross-entropy (BCE) loss is often used between the codeword and its estimate:
\begin{equation}
    \mathcal{L}(\textbf{x}^n, \hat{\textbf{x}}^n) = - \sum_{i=1}^{n} x_i \log \hat{x}_i + (1 - x_i) \log (1 - \hat{x}_i)  \ .
\end{equation}
Better results were obtained by averaging the BCE loss after each message passing iteration \cite{neural_bp1, neural_bp2}. However, reduction in the BCE loss does not necessarily result in BLER and BER, yet training the model directly using these metrics is not possible as they are not differentiable over the whole vector space $\mathbb{R}^n$. In the error correcting code transformer \cite{ECCT}, which exhibited promising results for a wide range of linear block codes, authors compute a loss between binary multiplicative noise and model output aiming at good noise prediction for a given code structure. In \cite{neuralbcjr}, the authors aim at obtaining a performance similar to the BCJR algorithm known to be optimal for convolutional codes. Thus, they use a mean squared error (MSE) loss between the model output and the result of the conventional BCJR decoder for training.

\subsection{Adversarial Attacks}

Adversarial attacks aim at decreasing the neural network performance by introducing minor perturbations into its input. Channel decoding can be considered as a classification task with $2^k$ classes, where the decoder estimates the class of the transmitted message ${m} \in \{0, 1\}^k$. Let us denote here a neural network input as $S$, the corresponding true label as $m$ and the loss function as $J(f(S), m)$, for example, the cross-entropy loss. A non-targeted adversarial attack is considered successful if the classifier misclassifies the adversarial sample: $f(S^{adv}) \neq m$, while the $p$-norm distance between the adversarial and original samples satisfies the constraint $||S^{adv} - S||_{p} < \epsilon$, as the goal is to make the attack as invisible to humans as possible. The optimization problem for generating an adversarial example is:
\begin{equation}
    S^{adv} = \argmax_{||S^{adv} - S||_{p} < \epsilon} J(f(S^{adv}), m) \ . \label{adv_problem}
\end{equation}
There are different attack methods which solve this optimization problem. The fast gradient sign method (FGSM) \cite{fgsm} solves (\ref{adv_problem}) for $p=\infty$:
\begin{equation}
    S^{adv} = S + \epsilon \cdot \sign (\nabla_S J(f(S), m)) \ . \label{fgsm}
\end{equation}
The iterative version of FGSM (I-FGSM) \cite{ifgsm} applies the gradient update in (\ref{fgsm}) multiple times with some rate $\alpha$:
\begin{equation}
    S^{adv}_0 = S, \ S^{adv}_{t} = S^{adv}_{t-1} + \alpha \sign (\nabla_S J(f(S^{adv}_{t-1}), m)). \label{ifgsm}
\end{equation}
Here, we can set $\alpha = \frac{\epsilon}{T}$, where $T$ is the number of iterations, for consistency with (\ref{adv_problem}). It has been shown that iterative methods achieve better results in terms of successful attacks \cite{scale_attack}. The attack in (\ref{fgsm}), (\ref{ifgsm}) can also be generalised to other $p$-norms.

\begin{algorithm}[tb]
\caption{Search for Friendly Attack}
\label{alg:algorithm}
\textbf{Input}: modulated codeword $S^N$, codeword or message $m$\\
\textbf{Parameters}: batch size $B$, number of iterations $I$, noise std $\sigma$, $gradient\_scheduler$ \\ 
\textbf{Output}: $\hat{m}$
\begin{algorithmic}[1] 
\STATE Let $i=0$,  $s_0=S^N$ repeated $B$ times, $y_{true} = m$ repeated $B$ times, $a = 0$.
\WHILE{$i < I$}
\STATE $\epsilon = gradient\_scheduler (i)$
\STATE $(z_i)_{jk}\sim \mathcal{N}(0, \sigma), 1\leq j\leq B, 1\leq k\leq N$
\STATE $y_i = s_i + z_i$, 
\STATE $l_i = demodulator (y_i)$ is a $B \times n$ matrix 
\STATE calculate $\BER$, $\BLER$ for $decoder(l_i), y_{true}$
\STATE $a_i = - \epsilon \nabla_{s_i} J(decoder(l_i), y_{true})$
\STATE $\overline{a_i} = \frac{1}{B}\sum_{j=1}^{B}{a_i}_{j}$ average over batch
\STATE $\overline{a_i} = \overline{a_i}$ repeated $B$ times
\STATE $\hat{l_i} = demodulator (s_i + \overline{a_i} + z_i)$ 
\STATE calculate $\hat{\BER}$, $\hat{\BLER}$ for $decoder(\hat{l_i}), y_{true}$
\IF { $\hat{\BER} < \BER$ and/or $\hat{\BLER} < \BLER$}
\STATE $i = i + 1$, $s_i = s_i + \overline{a_i}$, $a = a + \overline{a_i}$
\ENDIF
\ENDWHILE
\STATE \textbf{return} $a$
\end{algorithmic}
\end{algorithm}

\section{Friendly Attack Against a Channel Decoder}

In this work, we suggest a new concept called a `friendly attack' on channel decoders to improve their reliability.

\subsection{Modified Scheme}

As previously mentioned, the decoder can be considered as a classifier with $2^k$ classes as it recovers a message $\hat{m} \in \{0,1\}^k$. The idea is to use the gradient method to find a perturbation vector for each codeword that will improve the decoding performance, rather than attacking it. In our scheme, we suggest to add the perturbation vector $a^N$ to the modulated codeword ${S}^N$ before transmission over the channel as shown in Fig. \ref{fig:scheme_attack}. The aim is to find such a vector $a^N$ that without violation of the average power constraint the decoding error (BLER or BER) on average over the channel distribution will be smaller. So the model $f$ under attack is not only decoder itself but also the noisy channel and demodulator as illustrated with a blue box in Fig. \ref{fig:scheme_attack}.

The differences between a conventional adversarial setting and our formulation are as follows:
\begin{itemize}
    \item As we want to increase the performance of the decoder, the optimization problem should be modified to:
    \begin{equation}
        S^{frnd} = \argmin_{S^{frnd}} J(f(S^{frnd}), m) \ . \label{dec_problem}
    \end{equation}
    It can be considered as a similar formulation to the targeted attack \cite{ifgsm}, but here the target class is the true one.
    \item In the optimisation problem in (\ref{dec_problem}) there is no constraint on a $p$-norm distance between the codeword and the attacked version $S^{frnd}={S}^N + a^N$, i.e., there is no constraint on the $p$-norm of vector $a^N$. Nevertheless, the average power constraint has to be preserved for an attacked codeword ${S}^N + a^N$. This can be achieved by applying an additional normalisation before transmitting the codeword: $$||C({S}^N + a^N)||_2^2 \leq NP \ ,$$
    where $C = \frac{\sqrt{NP}}{||{S}^N + a^{N}||_2}$ is the normalisation constant. 
    Thus, we can rewrite (\ref{ifgsm}) into an iterative gradient attack:
    \begin{equation}
    S^{frnd}_0 = S, \ S^{frnd}_{t} = S^{frnd}_{t-1} - \epsilon \nabla_S J(f(S^{frnd}_{t-1}), m) \ , \label{my_ifgm}
    \end{equation}
    or in terms of the attack vector:
    \begin{equation}
    a^{N}_0 = 0, \ a^{N}_{t} = a^{N}_{t-1} - \epsilon \cdot \nabla_S J(f(S + a^{N}_{t-1}), m) \ . \label{my_ifgm_a}
    \end{equation}
    \item The model we intend to attack is stochastic due to the presence of channel noise, which constitutes the primary challenge in our task. We want the attack vector $a^N$ for the modulated codeword $S^N$ to be good on average for noisy transmission in terms of BLER and BER. 
\end{itemize}

In general, any given decoder splits the channel output space into $2^k$ decision regions and the input of the decoder is a vector of LLRs $L^n$ from one of these regions.
The modulated codewords then might be not optimal for the fixed decoder as they may not correspond to the center of the decision regions. Thus, a small shift of the modulated codeword in the direction of the decision region's center should increase the decoding performance. We will say that our friendly attack is successful if it improves the decoding performance on average for a given codeword. As we consider linear codes here, the successful attack vector for one codeword can be easily applied to all codewords (with appropriate sign adaptation) due to the code linearity and must show similar improvements. Let us suppose that we have found an attack $a^N$ for an all-zero codeword for some modulation with a constellation point corresponding to all zero bits $c_0$ ($c_0 \in \mathrm{R}$ for real constellations, $c_0 \in \mathrm{C}$ for complex constellations), then the attacked codeword will be:
\begin{equation}
    S^{frnd} = S^N + \frac{S^N a^N}{S_0} \ ,
\end{equation}
where normalised attack vector $\frac{S^N a^N}{S_0} = \frac{1}{S_0}\{S_i a_i\}_{i=1}^{N}$.

\begin{figure}[t]
\centering
\includegraphics[width=0.9\columnwidth]{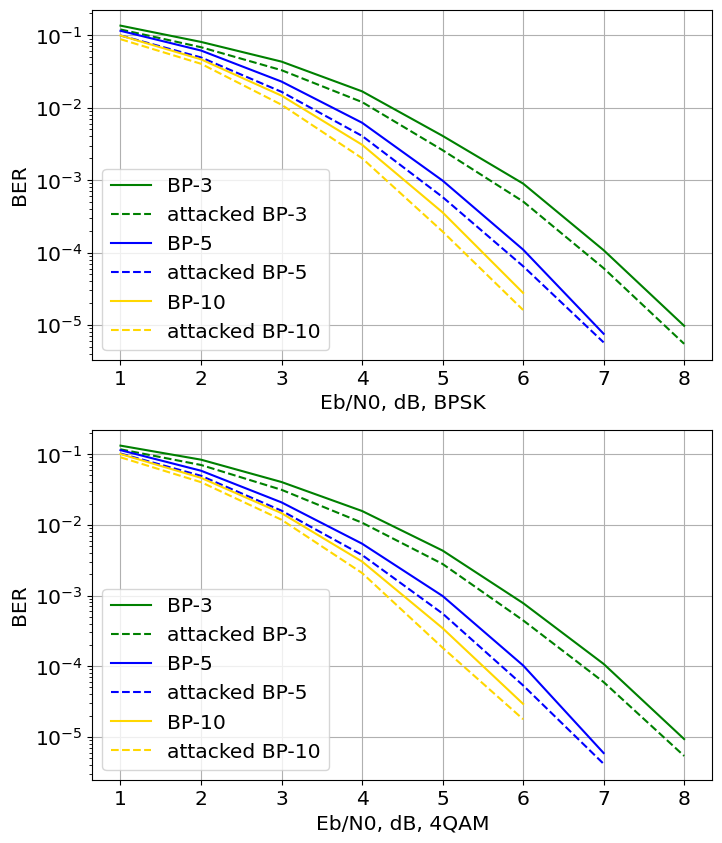}
\caption{BER for LDPC code $(64, 32)$ with and without attack using BPSK (top) and $4$-QAM (bottom) modulations. `BP-$n$' refers to BP decoder with $n$ iterations.}
\label{ldpc_64}
\end{figure}

\subsection{Attack}

To design an attack vector $a^N$ we need to be able to compute $\nabla_S J(f(S), m)$. The model $f$ consists of the noisy channel, the demodulator and the channel decoder as in Fig. \ref{fig:scheme_attack}. AWGN channel and the demodulator are differentiable. 
We consider BP decoder as it is differentiable, as well as NBP and other neural decoders such as ECCT \cite{ECCT} for linear codes.

To overcome the problem with stochasticity of the transmission we apply a gradient update (\ref{my_ifgm_a}) over a batch of size $B$ as in Alg. \ref{alg:algorithm}. That means we send the same codeword $S^N$ over the channel $B$ times for every iteration (line $3$ in Alg. \ref{alg:algorithm}). In this section and in Alg. \ref{alg:algorithm}, we use $\nabla_S J(f(S), m)$ as a gradient from the decoder for the batch of codewords $S^N$. All $B$ attack vectors are then averaged over the batch and added to the modulated codeword to check (line $11$) if they indeed improve the decoding performance. This is not necessary for all architectures we tried, but crucial for architectures like ECCT as it aims at noise prediction, and the gradient direction obtained for this decoder might be bad on average over the batch.

\begin{figure*}[t]
\centering
\includegraphics[width=1.8\columnwidth]{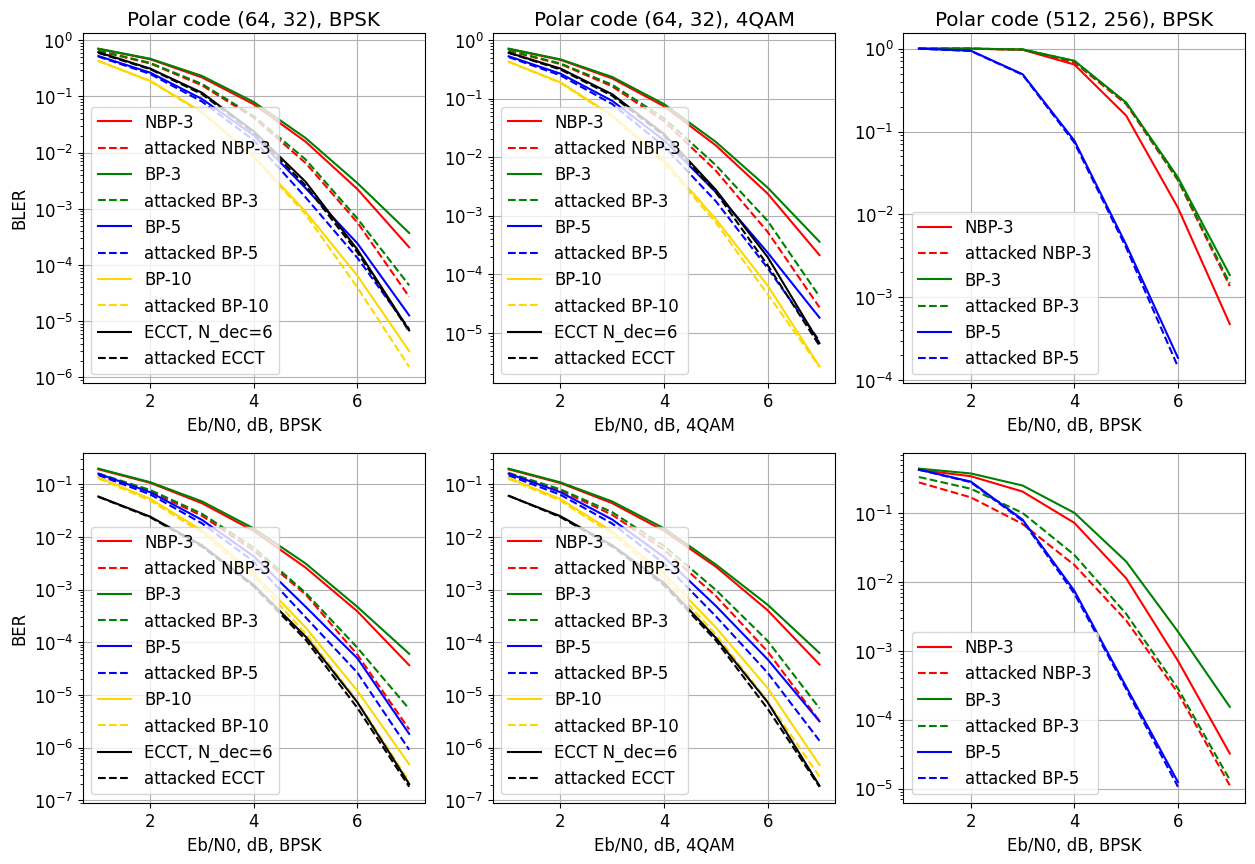} 
\caption{BLER (top) and BER (bottom) for $(64, 32)$ and $(512, 256)$ polar codes with BPSK and $4$-QAM modulations.}
\label{polar}
\end{figure*}

The choice of parameters $I, B$, and a scheduler for $\epsilon$ defines an attack. We can highlight four different approaches:
\begin{enumerate}
    \item In the first approach, we employ a rather large batch size $B \sim 1000-10000$ and relatively small number of iterations $I \sim 1-100$.
    \item The second approach needs an average batch size $B \sim 100-500$ and relatively high number of iterations $I \sim 500-5000$. For this setting a scheduler for $\epsilon$ should be chosen carefully. If this approach is applicable, it gives the best overall results. However, it does not perform well for some decoders and modulations due to a big bias towards the noise in a batch.
    \item The third and fourth approaches are more robust as they need multiple runs of Alg. \ref{alg:algorithm} and saving output attack vectors after each run. 
    The final attack vector is found by running a clustering algorithm on the set of attack vectors from multiple runs and then choosing the best one. The best results were obtained with Agglomerative Clustering (with linkage criterion "ward" and "complete") and K-means (with number of neighbours $3-4$). For both approaches, the scikit-learn library implementation \cite{scikit-learn} was used. Finally, the third approach includes a relatively small batch size $B \sim 5-50$, number of iterations $I \sim 15-40$ and many runs of algorithm \ref{alg:algorithm}: $1000-5000$.
    \item The fourth approach is similar to the first one but is run multiple times with a small number of iterations $I \sim 1-5$. This method is time consuming and works for neural models which were trained on the input $\pm1$ with  little noise introduced as they are vulnerable to any perturbation of the input.
\end{enumerate}

Such a variety of approaches in attacks and their different results in different settings (code, decoder type, modulation) can be explained by the different splits of the codeword space. If there is more than one direction for the gradient descent the algorithm may struggle, and the third and fourth approaches work the best because final clustering allows to have more than one optimal perturbation vector.

Taking the gradient $\nabla_S J(f(S), m)$ during decoding and adding it to a codeword can be interpreted as greater power allocation for symbol positions in $S^N$ which have more impact on the error correction. The decoder corrects errors only if there is some noise introduced during transmission otherwise there is no error to correct. It means that it is possible to find a successful attack vector only during noisy transmission. If an attack vector is obtained in a high signal-to-noise ratio (SNR) regime (almost no noise), the decoder has almost no errors to correct, and it is difficult to identify an effective attack vector. If an attack vector is obtained at a low SNR (SNRs for which $\mathrm{BLER}\geq0.7$), there is also almost no error correction because the decoder is not able to extract any information from the noisy channel output. We saw in our experiments that the best attack vectors are found in SNR regimes where BLER $\sim ~0.1-0.5$.

\begin{figure}[t]
\centering
\includegraphics[width=0.92\columnwidth]{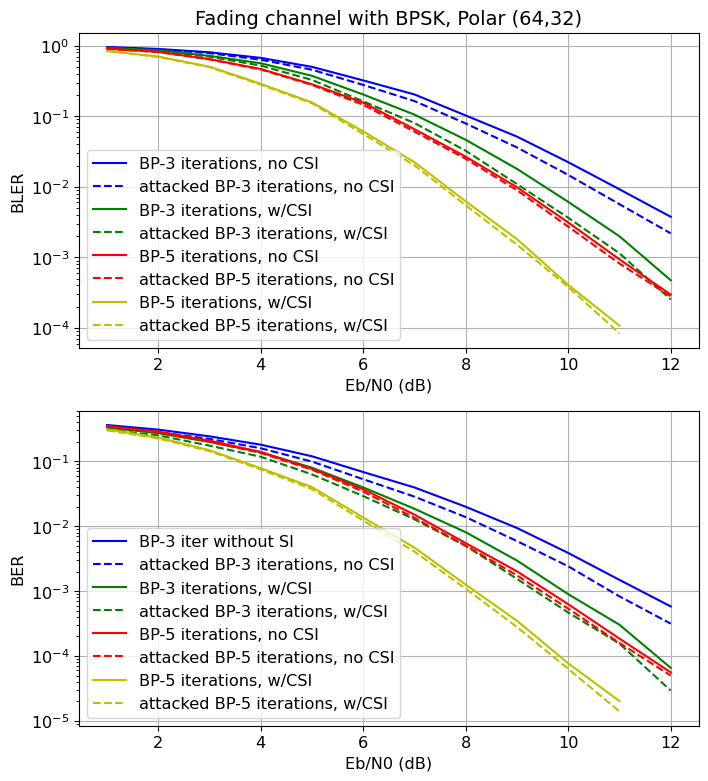}
\caption{BLER (top) and BER (bottom) for $(64, 32)$ polar code over a fading channel with BPSK.}
\label{polar_fading}
\end{figure}

\section{Simulation Results}

In this work, we do not aim at proposing a new efficient decoding approach, but to show that our friendly attack can improve the performance of the variety of existing codes and decoding algorithms in different scenarios. To compare our results with the state-of-the-art approaches we used the sionna library \cite{sionna}, which follows the 5G standard \cite{3gpp}. We consider both polar and LDPC codes with different decoding approaches and modulations. LDPC code design in our experiments is the same as in the sionna library. We are particularly interested in short block lengths as near capacity performance can be achieved by LDPC codes in large block lengths. 
All attacks were obtained using the all-zero codeword, and then tested on random codewords. All the simulations were run on Nvidia RTX A6000 GPUs with 48GB of memory.


We first consider the $(64, 32)$ LDPC code with BP decoding \cite{LDPC_shannon} for different number of iterations and different modulations. BER of BP decoding with $3$ and $5$ iterations and their best attacks are presented in Fig. \ref{ldpc_64} for BPSK and $4$-QAM modulations.  Improvement for BLER performance for these decoders via friendly attacks has similar behaviour and is omitted here. We consider here relatively small number of iterations, since with the increasing number of iterations, the BP decoder gets closer to the optimal decoding, and is harder to improve.

For comparison we considered also polar codes: $(64, 32)$ for a short block length and $(512, 256)$ for a long block length. We consider both BP and NBP decoding for polar codes. The results are presented in Fig. \ref{polar}. For long block length, $k=256$, there is no improvement in BLER for NBP while BER performance was noticeably enhanced via the friendly attack for both BP and NBP decoders. It can be noticed that improvement in BER for BP decoding is greater than that for NBP. We also include the ECCT \cite{ECCT} decoding results with $N_{dec}=6$ for code $(64, 32)$ and its attacks for both modulations. Although the improvement via attack for ECCT is barely visible, it shows around \%10 reduction in BER and BLER for $>4$ dB even for this complex and deep model. 

In Fig. \ref{polar_fading}, we present BLER and BER results for polar code $(64, 32)$ over a Rayleigh fading channel with BPSK modulation. For this scenario, we considered BP decoding with $3$ and $5$ iterations both with and without side information (SI). Similarly to previous results, more significant improvement can be achieved for the suboptimal BP decoder with 3 iterations. 

\begin{figure}[t]
\centering
\includegraphics[width=0.8\columnwidth]{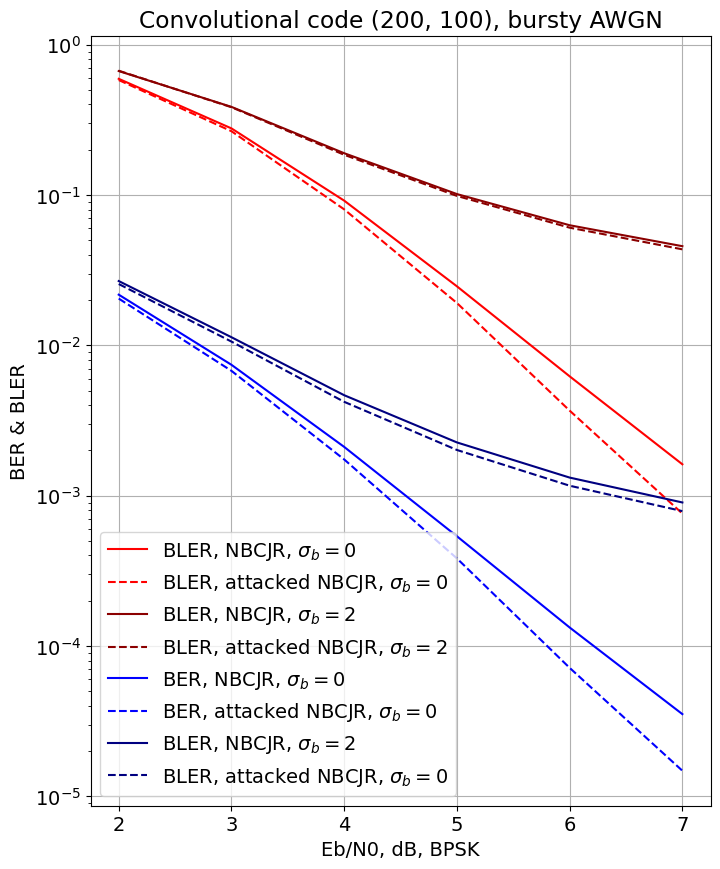}
\caption{BLER and BER for the $(200, 100)$ convolutional code with BPSK modulation and NBCJR decoding.}
\label{convcode_bcjr}
\end{figure}

As an additional evidence of the adaptability of the friendly attack method we also provide the results for NBCJR \cite{neuralbcjr} for convolutional code of rate $R=\frac{1}{2}$ and block length $k=100$ in Fig. \ref{convcode_bcjr} for both AWGN and bursty AWGN channels.

\section{Conclusion}

In this work, we proposed a new concept of a 'friendly attack' for channel coding that employs the adversarial attack approach to enhance the channel decoding. The friendly attack is a vector found by a proposed here iterative algorithm based on a gradient descent during transmission and decoding. We showed that the addition of the attack vector to a modulated codeword before transmission significantly improves the error correction performance of the decoder without any changes on the decoder's side. The proposed approach was tested for a range of codes and decoders, and also for different modulations and for different channels. Proposed friendly attack showed promising results in terms of BER for different decoders and it appeared to be more efficient in BER improvement for BP decoding compared to NBP for a small number of decoding iterations. 

\section{Acknowledgements}
This work received funding from the UKRI for the project AIR (European Resarch Council Consolidator Grant, EP/X030806/1). 

\bibliography{aaai24}

\begin{thebibliography}{36}
\providecommand{\natexlab}[1]{#1}

\bibitem[{{3GPP}(2018)}]{3gpp}
{3GPP}. 2018.
\newblock Multiplexing and channel coding (Release 10) 3GPP TS 21.101 v10.4.0.
\newblock \url{https://www.3gpp.org/ftp/Specs/2023-06/Rel-10/21_series/21101-a40.zip}.
\newblock Accessed: 2023-08-13.

\bibitem[{Aref and Chagnon(2022)}]{Aref:OFC:22}
Aref, V.; and Chagnon, M. 2022.
\newblock End-to-End Learning of Joint Geometric and Probabilistic Constellation Shaping.
\newblock In \emph{2022 Optical Fiber Communications Conference and Exhibition (OFC)}, 1--3.

\bibitem[{Arikan(2008)}]{polar}
Arikan, E. 2008.
\newblock Channel polarization: A method for constructing capacity-achieving codes.
\newblock In \emph{2008 IEEE International Symposium on Information Theory}, 1173--1177.

\bibitem[{Bennatan, Choukroun, and Kisilev(2018)}]{syndrom_based}
Bennatan, A.; Choukroun, Y.; and Kisilev, P. 2018.
\newblock Deep Learning for Decoding of Linear Codes - A Syndrome-Based Approach.

\bibitem[{Berrou, Glavieux, and Thitimajshima(1993)}]{turbo}
Berrou, C.; Glavieux, A.; and Thitimajshima, P. 1993.
\newblock Near Shannon limit error-correcting coding and decoding: Turbo-codes. 1.
\newblock In \emph{Proceedings of ICC '93 - IEEE International Conference on Communications}, volume~2, 1064--1070 vol.2.

\bibitem[{Buchberger et~al.(2020)Buchberger, Häger, Pfister, Schmalen, and Graell~i Amat}]{neural_bp_impr}
Buchberger, A.; Häger, C.; Pfister, H.; Schmalen, L.; and Graell~i Amat, A. 2020.
\newblock Learned Decimation for Neural Belief Propagation Decoders.

\bibitem[{Choukroun and Wolf(2022)}]{ECCT}
Choukroun, Y.; and Wolf, L. 2022.
\newblock Error Correction Code Transformer.
\newblock arXiv:2203.14966.

\bibitem[{Doan, Ali~Hashemi, and Gross(2018)}]{neural_sc}
Doan, N.; Ali~Hashemi, S.; and Gross, W.~J. 2018.
\newblock Neural Successive Cancellation Decoding of Polar Codes.
\newblock In \emph{2018 IEEE 19th International Workshop on Signal Processing Advances in Wireless Communications (SPAWC)}, 1--5.

\bibitem[{Doan et~al.(2018{\natexlab{a}})Doan, Hashemi, Mambou, Tonnellier, and Gross}]{neural_bp_polar_crc}
Doan, N.; Hashemi, S.~A.; Mambou, E.~N.; Tonnellier, T.; and Gross, W.~J. 2018{\natexlab{a}}.
\newblock Neural Belief Propagation Decoding of CRC-Polar Concatenated Codes.
\newblock \emph{ICC 2019 - 2019 IEEE International Conference on Communications (ICC)}, 1--6.

\bibitem[{Doan et~al.(2018{\natexlab{b}})Doan, Hashemi, Mondelli, and Gross}]{bp_polar_permuted1}
Doan, N.; Hashemi, S.~A.; Mondelli, M.; and Gross, W. 2018{\natexlab{b}}.
\newblock On the Decoding of Polar Codes on Permuted Factor Graphs.
\newblock 1--6.

\bibitem[{Elias(1955)}]{elias}
Elias, P. 1955.
\newblock Coding for noisy channels.
\newblock \emph{IRE Conv. Rec.}, 3: 37--46.

\bibitem[{Elkelesh et~al.(2018)Elkelesh, Ebada, Cammerer, and ten Brink}]{bp_polar_permuted2}
Elkelesh, A.; Ebada, M.; Cammerer, S.; and ten Brink, S. 2018.
\newblock Belief propagation decoding of polar codes on permuted factor graphs.
\newblock In \emph{2018 IEEE Wireless Communications and Networking Conference (WCNC)}, 1--6.

\bibitem[{Gallager(1962)}]{LDPC}
Gallager, R. 1962.
\newblock Low-density parity-check codes.
\newblock \emph{IRE Transactions on Information Theory}, 8(1): 21--28.

\bibitem[{Goodfellow, Shlens, and Szegedy(2015)}]{fgsm}
Goodfellow, I.~J.; Shlens, J.; and Szegedy, C. 2015.
\newblock Explaining and Harnessing Adversarial Examples.
\newblock In Bengio, Y.; and LeCun, Y., eds., \emph{International Conference on Learning Representations, {ICLR} 2015, San Diego, CA, USA, May 7-9, 2015, Conference Track Proceedings}.

\bibitem[{Gruber et~al.(2017)Gruber, Cammerer, Hoydis, and Brink}]{on_deeplearning_based_cc}
Gruber, T.; Cammerer, S.; Hoydis, J.; and Brink, S.~t. 2017.
\newblock On deep learning-based channel decoding.
\newblock In \emph{2017 51st Annual Conference on Information Sciences and Systems (CISS)}, 1--6.

\bibitem[{Gümüş et~al.(2020)Gümüş, Alvarado, Chen, Häger, and Agrell}]{Gumus:OFC:20}
Gümüş, K.; Alvarado, A.; Chen, B.; Häger, C.; and Agrell, E. 2020.
\newblock End-to-End Learning of Geometrical Shaping Maximizing Generalized Mutual Information.
\newblock In \emph{2020 Optical Fiber Communications Conference and Exhibition (OFC)}, 1--3.

\bibitem[{Hameed, György, and Gündüz(2021)}]{Hameed:TIFS:21}
Hameed, M.~Z.; György, A.; and Gündüz, D. 2021.
\newblock The Best Defense Is a Good Offense: Adversarial Attacks to Avoid Modulation Detection.
\newblock \emph{IEEE Transactions on Information Forensics and Security}, 16: 1074--1087.

\bibitem[{Hou, Siegel, and Milstein(2001)}]{rayleigh}
Hou, J.; Siegel, P.; and Milstein, L. 2001.
\newblock Performance analysis and code optimization of low density parity-check codes on Rayleigh fading channels.
\newblock \emph{IEEE Journal on Selected Areas in Communications}, 19(5): 924--934.

\bibitem[{Hoydis et~al.(2022)Hoydis, Cammerer, {Ait Aoudia}, Vem, Binder, Marcus, and Keller}]{sionna}
Hoydis, J.; Cammerer, S.; {Ait Aoudia}, F.; Vem, A.; Binder, N.; Marcus, G.; and Keller, A. 2022.
\newblock Sionna: An Open-Source Library for Next-Generation Physical Layer Research.
\newblock \emph{arXiv preprint}.

\bibitem[{Jones, Yankov, and Zibar(2019)}]{Jones:ECOC:19}
Jones, R.~T.; Yankov, M.~P.; and Zibar, D. 2019.
\newblock End-to-end learning for GMI optimized geometric constellation shape.
\newblock In \emph{45th European Conference on Optical Communication (ECOC 2019)}, 1--4.

\bibitem[{Kim et~al.(2018)Kim, Jiang, Rana, Kannan, Oh, and Viswanath}]{neuralbcjr}
Kim, H.; Jiang, Y.; Rana, R.~B.; Kannan, S.; Oh, S.; and Viswanath, P. 2018.
\newblock Communication Algorithms via Deep Learning.
\newblock In \emph{International Conference on Learning Representations}.

\bibitem[{Kurakin, Goodfellow, and Bengio(2016)}]{ifgsm}
Kurakin, A.; Goodfellow, I.; and Bengio, S. 2016.
\newblock Adversarial examples in the physical world.

\bibitem[{Kurakin, Goodfellow, and Bengio(2017)}]{scale_attack}
Kurakin, A.; Goodfellow, I.~J.; and Bengio, S. 2017.
\newblock Adversarial Machine Learning at Scale.
\newblock In \emph{5th International Conference on Learning Representations, {ICLR} 2017, Toulon, France, April 24-26, 2017, Conference Track Proceedings}. OpenReview.net.

\bibitem[{Lugosch and Gross(2017)}]{neural_minsum}
Lugosch, L.; and Gross, W.~J. 2017.
\newblock Neural Offset Min-Sum Decoding.
\newblock In \emph{2017 IEEE International Symposium on Information Theory (ISIT)}, 1361–1365. IEEE Press.

\bibitem[{MacKay and Neal(1997)}]{LDPC_shannon}
MacKay, D.~J.; and Neal, R.~M. 1997.
\newblock Near Shannon limit performance of low density parity check codes.
\newblock volume~33, 457--458.

\bibitem[{Makkuva et~al.(2021)Makkuva, Liu, Jamali, Mahdavifar, Oh, and Viswanath}]{KOcodes}
Makkuva, A.~V.; Liu, X.; Jamali, M.~V.; Mahdavifar, H.; Oh, S.; and Viswanath, P. 2021.
\newblock KO codes: inventing nonlinear encoding and decoding for reliable wireless communication via deep-learning.
\newblock In Meila, M.; and Zhang, T., eds., \emph{Proceedings of the 38th International Conference on Machine Learning}, volume 139 of \emph{Proceedings of Machine Learning Research}, 7368--7378. PMLR.

\bibitem[{Nachmani, Be'ery, and Burshtein(2016)}]{neural_bp1}
Nachmani, E.; Be'ery, Y.; and Burshtein, D. 2016.
\newblock Learning to decode linear codes using deep learning.
\newblock In \emph{2016 54th Annual Allerton Conference on Communication, Control, and Computing (Allerton)}, 341--346.

\bibitem[{Nachmani et~al.(2018)Nachmani, Marciano, Lugosch, Gross, Burshtein, and Be’ery}]{neural_bp2}
Nachmani, E.; Marciano, E.; Lugosch, L.; Gross, W.~J.; Burshtein, D.; and Be’ery, Y. 2018.
\newblock Deep Learning Methods for Improved Decoding of Linear Codes.
\newblock \emph{IEEE Journal of Selected Topics in Signal Processing}, 12(1): 119--131.

\bibitem[{O'Shea and Hoydis(2017)}]{intro_to_ml_cc}
O'Shea, T.; and Hoydis, J. 2017.
\newblock An Introduction to Machine Learning Communications Systems.
\newblock \emph{ArXiv}, abs/1702.00832.

\bibitem[{Pedregosa et~al.(2011)Pedregosa, Varoquaux, Gramfort, Michel, Thirion, Grisel, Blondel, Prettenhofer, Weiss, Dubourg, Vanderplas, Passos, Cournapeau, Brucher, Perrot, and Duchesnay}]{scikit-learn}
Pedregosa, F.; Varoquaux, G.; Gramfort, A.; Michel, V.; Thirion, B.; Grisel, O.; Blondel, M.; Prettenhofer, P.; Weiss, R.; Dubourg, V.; Vanderplas, J.; Passos, A.; Cournapeau, D.; Brucher, M.; Perrot, M.; and Duchesnay, E. 2011.
\newblock Scikit-learn: Machine Learning in {P}ython.
\newblock \emph{Journal of Machine Learning Research}, 12: 2825--2830.

\bibitem[{Polyanskiy, Poor, and Verd\'{u}(2010)}]{bounds_polyanskiy}
Polyanskiy, Y.; Poor, H.~V.; and Verd\'{u}, S. 2010.
\newblock Channel Coding Rate in the Finite Blocklength Regime.
\newblock 56(5): 2307–2359.

\bibitem[{Qu and Djordjevic(2019)}]{prob_geom_shaping}
Qu, Z.; and Djordjevic, I.~B. 2019.
\newblock On the Probabilistic Shaping and Geometric Shaping in Optical Communication Systems.
\newblock \emph{IEEE Access}, 7: 21454--21464.

\bibitem[{Ren et~al.(2015)Ren, Zhang, Liu, and You}]{bp_polar_earlystop}
Ren, Y.; Zhang, C.; Liu, X.; and You, X. 2015.
\newblock Efficient early termination schemes for belief-propagation decoding of polar codes.
\newblock In \emph{2015 IEEE 11th International Conference on ASIC (ASICON)}, 1--4.

\bibitem[{Shannon(1948)}]{shannon_capacity}
Shannon, C.~E. 1948.
\newblock A mathematical theory of communication.
\newblock \emph{The Bell System Technical Journal}, 27(3): 379--423.

\bibitem[{Tal and Vardy(2011)}]{scl_polar}
Tal, I.; and Vardy, A. 2011.
\newblock List decoding of polar codes.
\newblock In \emph{2011 IEEE International Symposium on Information Theory Proceedings}, 1--5.

\bibitem[{Xu et~al.(2017)Xu, Wu, Ueng, You, and Zhang}]{neural_bp_polar}
Xu, W.; Wu, Z.; Ueng, Y.-L.; You, X.; and Zhang, C. 2017.
\newblock Improved polar decoder based on deep learning.
\newblock In \emph{2017 IEEE International Workshop on Signal Processing Systems (SiPS)}, 1--6.

\end{thebibliography}

\end{document}